\documentclass[twocolumn,prl,amsmath,amssymb,amsfonts]{revtex4-1}
\usepackage{graphicx}
\usepackage{color}
\usepackage[breaklinks=false]{hyperref}

\newcommand{\coli}{\textit{E. coli} }

\begin{document}

\title{Polar features in the flagellar propulsion of  \coli bacteria}

\author{S. Bianchi$^1$}
\author{F. Saglimbeni$^1$}
\author{A. Lepore$^1$}
\author{R. Di Leonardo$^{1,2}$}
\email{roberto.dileonardo@phys.uniroma1.it}

\affiliation{ $^1$Dipartimento di Fisica, Universit\`a di Roma ``Sapienza'',
I-00185, Roma, Italy }
\affiliation{ $^2$CNR-NANOTEC, Soft and Living Matter Laboratory,  I-00185 Roma, Italy}
\date{\today}

\begin{@twocolumnfalse}
\begin{center}
\mbox{
{
\color{blue} \href{http://dx.doi.org/10.1103/PhysRevE.91.062705}{http://dx.doi.org/10.1103/PhysRevE.91.062705}
}
}
\end{center}
\end{@twocolumnfalse}

\begin{abstract}
\coli bacteria swim following a run and tumble pattern. In the run state all
flagella join in a single helical bundle that propels the cell body along
approximately straight paths. When one or more flagellar motors reverse
direction the bundle unwinds and the cell randomizes its orientation. This basic
picture represents an idealization of a much more complex dynamical problem.
Although it has been shown that bundle formation can occur at either pole of the
cell, it is still unclear whether this two run states correspond to asymmetric
propulsion features. Using holographic microscopy we record the 3D motions of
individual bacteria swimming in optical traps. We find that  most cells  possess
two run states characterised by different propulsion forces, total torque and
bundle conformations. We analyse the statistical properties of bundle reversal
and compare the hydrodynamic features of forward and backward running states.
Our method is naturally multi-particle and opens up the way towards controlled
hydrodynamic studies of interacting swimming cells.

\end{abstract}

\maketitle

\section{Introduction}

\textit{Escherichia coli} is a peritrichously flagellated rod-shaped bacterium
with no apparent signs of polarity.
It swims at about 20 $\mu$m/s following a characteristic run and tumble
pattern \cite{colibook}. When all flagella rotate in the same direction they
join together near a cell pole forming a compact  helical bundle that propels
the cell body along approximately straight runs. The bundle, however, has a
finite lifetime of about 1 s after which it disentangles due to one or more
flagellar motor changing direction \cite{Berg2000imaging}. In this transient
state, called tumble, the uncoordinated motions of flagella cause a rapid and
random reorientation of the cell body axis. After about 0.1 s all flagella
resume to the same rotation direction and a new bundle forms.
In a perfectly symmetric cell, bundle formation should occur with equal
probability at either cell pole and result in equal propulsion forces.
Conversely, polar flagellated bacteria like {\it P. putida}  have been observed
to swim  in opposite directions with very different speeds
\cite{Beta2013twospeeds}. However, even in the case of an apparently symmetric
cell like \coli, division by binary fission allows to distinguish between a new
and an old pole. An age asymmetry between the two poles will naturally give rise
to a corresponding asymmetry in all pole properties that depend on the cell age.
For example, the chemoreceptor Tsr has been shown to be more abundant at the old
pole \cite{ping2}. An accurate analysis of flagella distribution over the cell
surface also shows an increased density of flagella over the old pole cell half
\cite{ping}. In addition, since the average number of flagella is typically only
about 6 \cite{ping}, large deviations from a symmetric flagellar distribution are expected
to occur just by random chance.                   
These observations lead naturally to the question whether the motility properties
of \coli display any corresponding asymmetries, what is their nature and entity
and if there's any particular advantage that an asymmetric swimming could offer.
One of the first polar features that was recognised in the run and tumble motion
of \coli was that a large number of cells show a marked tendency of forming the
bundle at one specific pole after a tumble \cite{Turner1995reversal}. Later work
showed that the old pole is usually the preferred pole for bundle formation and
attributed this to a larger average number of flagella over the old pole cell half
\cite{ping}. Long term observation of tumbling \coli in a double optical trap
confirmed the presence of a preferred pole for bundle formation in the majority
of cells \cite{Chemla2009trapping}. The latter work also reported the presence
of two distinct body rotation frequencies corresponding to a bundle forming at
one pole or the other. However, the presence of two corresponding distinct
propelling forces has never been reported so far.

 A direct measurement of the propelling force is feasible in principle using
calibrated optical traps \cite{trapping}. Using a two trap setup, combined with
back focal plane detection of cross-polarised transmitted beams, is a common and
highly accurate way for measuring forces between two trapped spherical objects
\cite{dual}. Although a two trap configuration has been successfully applied for
trapping and alignment of rod-shaped bacteria \cite{Chemla2009trapping}, force
measurements are not straightforward when trapping a single elongated body. An
alternative approach consists in trapping a cell near the coverslip with a
single beam while simultaneously applying an external flow to align the cell on
the focal plane and gauge propelling forces with viscous drag
\cite{Chattopadhyay2006trapping}. The presence of the flow however could
strongly bias bundle formation towards one pole. Moreover, due to the proximity
of a solid wall, hydrodynamic wall effects can strongly affect force
measurements and swimming efficiency.

Here we demonstrate that a combination of single beam optical trapping and 3D
holographic microscopy provides an ideal tool to investigate in the bulk fluid
the propulsion features of the two possible run states together with a
careful examination of the statistics of transitions between them. We find that
the majority of cells displays two distinct run states depending on which side
the bundle forms. The average difference between the two corresponding
propelling forces is 35\% with the largest deviations exceeding 100\%. 

\section{Methods}

\begin{figure}[ht]
\includegraphics[width=0.5\textwidth]{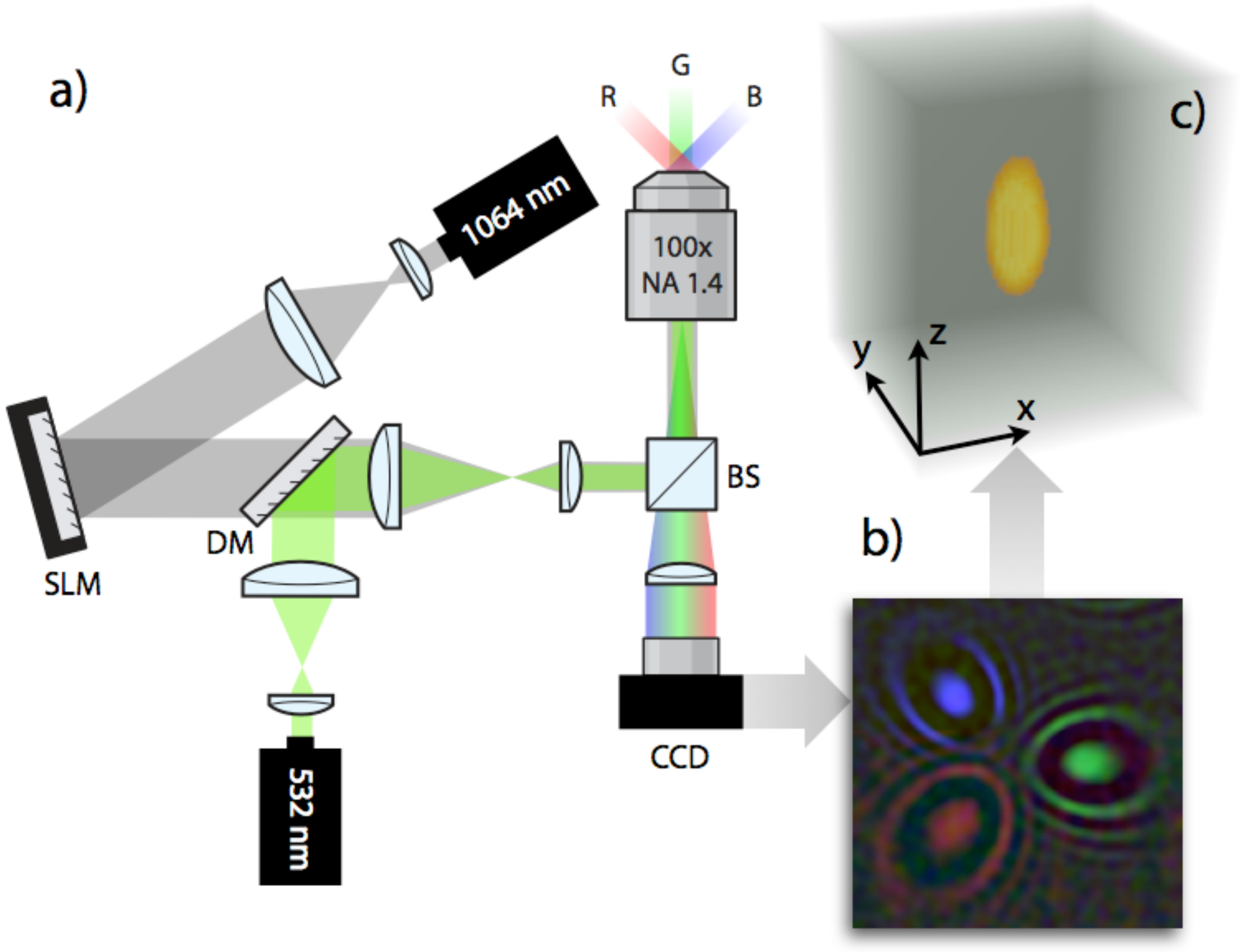}
\caption{(a) Optical setup: three RGB LED sources are used for 3-axes DHM
volumetric reconstructions of individually trapped cells. The trapping IR laser
(shown in gray) is phase modulated by a SLM and sent to the entrance pupil of a
100X, NA 1.45 oil immersion objective by a beam splitter (BS). A second 532 nm
laser is coupled to the objective and used to stop bacterial motility by
irreversible photodamage. (b) A typical RGB fringe pattern produced by a trapped
cell located 7 $\mu$m above the focal plane. (c) Corresponding 3D
reconstruction of the cell body.}
\label{fig:setup}
\end{figure}

\coli cells (MG1655) were streaked on a Petri dish containing 1.5\% of agar and 
tryptone broth (1\% tryptone and 0.5\% NaCl). A single-colony was inoculated
into tryptone broth and grown overnight at 32$^\circ$C. The saturated culture
was then diluted 1:100 (50 $\mu$l in 5 ml) into fresh medium and grown for 4
hours at 32$^\circ$C with 200 rpm rotation. Bacterial cells were then harvested
from culture media by centrifugation at 2200 rpm for 10 minutes at room
temperature. The pellet was resuspended by gently mixing in trapping buffer
composed of 100 mM Tris-HCl, 70 mM NaCl, 2\% (wt/vol) glucose and an oxygen
scavenging system \cite{Chemla2009trapping}. The cells were washed three times
to replace growth medium with trapping buffer. Trapping buffer sustain bacteria
motility and reduce the oxidative damage to the cell induced by trapping.
Individual cells were trapped using an infrared ($\lambda$=1064 nm) holographic
trapping system \cite{holo} based on a LCOS Spatial Light Modulator (Hamamatsu
X10468) Figure \ref{fig:setup}(a). Using an oxygen scavenging system trapped
bacteria can display a high motility for hours. 3D volumetric reconstructions of trapped bacteria are obtained using
a 3-axes/3-colours modification of in-line Digital Holographic Microscopy (DHM)
\cite{3axis}. In conventional in-line DHM the sample volume is reconstructed by
numerically back-propagating the complex field encoded in the interference
pattern between scattered and incident light \cite{dhm}. In our setup we
replaced the single on-axis illumination beam with three LED collimated light
beams of different wavelengths (Prizmatix UHP-Mic-LED-630, UHP-Mic-LED-520,
UHP-Mic-LED-450) chosen to match the spectral response of the three colour
channels of a RGB colour camera (Basler avA1000-100gm). The sample scatters the
light of the three beams producing three independent fringe patterns. The field
intensities corresponding to the three colour channels are reconstructed in 3D and
then are overlapped to obtain a volumetric image. Figure \ref{fig:setup}(b-c)
shows a RGB hologram with the corresponding 3D reconstruction of a trapped \coli. 
As in standard DHM, good numerical reconstructions are obtained when the scattering object 
is located a few microns above the focal plane \cite{3axis,dhm}.
To meet this requirement, we use the SLM to move the trapped cells 7 $\mu$m above the focal plane.

\section{Results}

We trapped individually 53 cells and observed their 3D motion at 100 frames per second.
Total observation time for each swimming cell was 5 minutes during which no
motility reduction was observed. Cells were trapped 40 $\mu$m above the
coverslip so that hydrodynamic coupling to the glass surface was negligible.

\begin{figure}[ht]
\includegraphics[width=0.475\textwidth]{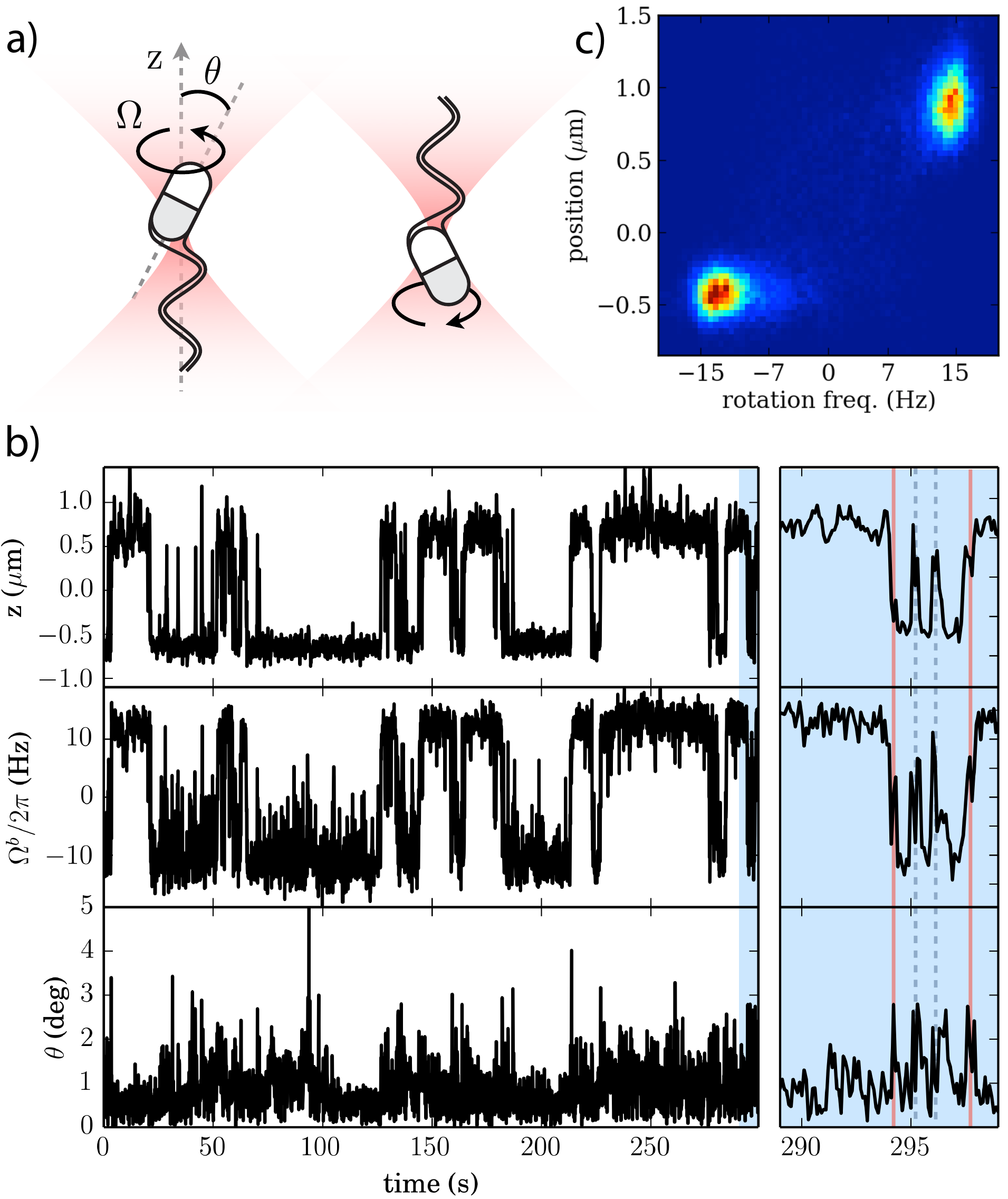}
\caption{(a) A trapped cell can form a bundle on either of its two poles. The
rotating bundle at one pole pushes the cell body towards the opposite pole. The
two swimming states also give rise to opposite cell body rotations around $z$.
(b) Time traces of the axial cell body displacement from the trap equilibrium
position $z$, rotational frequency $\Omega^b$, and cell body polar angle $\theta$. (c) 2D histogram obtained from
the $z$, $\Omega^b$ traces shown in (b).}
\label{fig:single}
\end{figure}

Rod shaped objects like \coli spontaneously align along the beam axis maximising
the overlap between the prolate cell body and the elongated focal distribution
of light. In addition to trapping gradient forces, that oppose cell
displacements away from the trap center, gradient torques arise restoring the
cell axis onto the beam axis.

\begin{figure}[ht]
\includegraphics[width=0.45\textwidth]{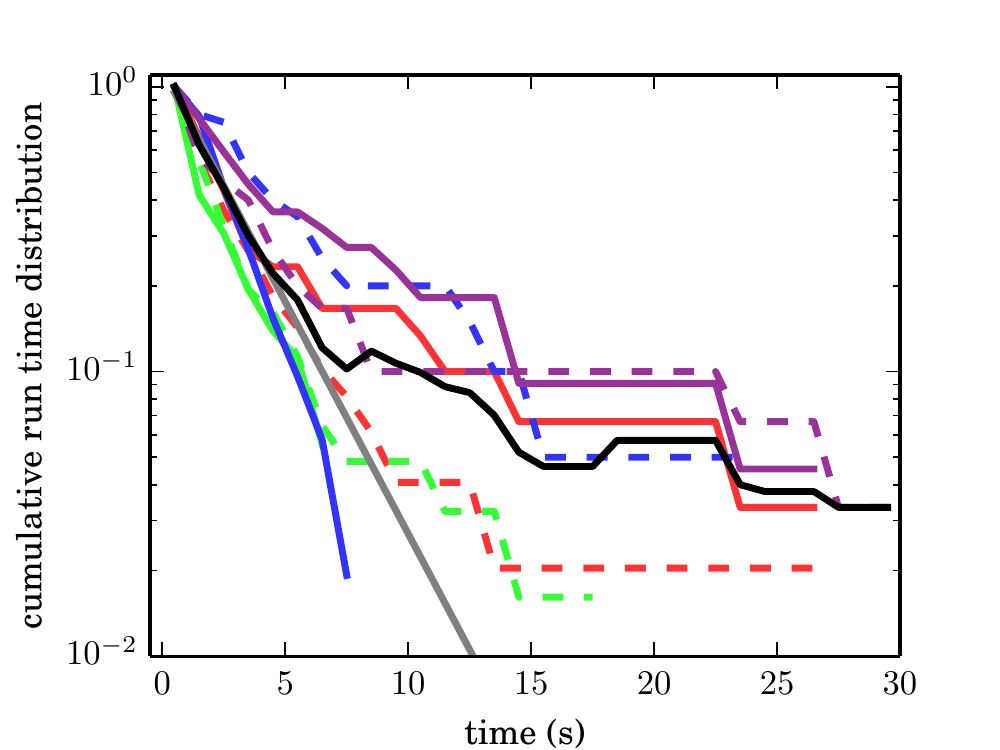}
\caption{Cumulative run length distributions. Different colours represent different cells. Solid and dashed lines distinguish between the two possible run states corresponding to having the flagellar bundle at each of the two cell poles. Black line is the logarithmic average of all the reported curves whose corresponding exponential fit is shown in gray.}
\label{fig:tau}
\end{figure}

During the run state, a flagellar bundle is formed on one pole pushing the cell
body away from the trap equilibrium axial position $z=0$. 
Due to a slight misalignment between the bundle and body axes, the cell body
precesses around $z$. The time evolution of cell position and orientation were
extracted from the first and second spatial moment of volumetric images. 
Figure \ref{fig:single}(b) shows typical time traces obtained for the axial
components of cell body position z, angular velocity $\Omega^b$  and polar angle $\theta$ of the body
axis. Synchronous transitions between two states are clearly visible in the $z$
and $\Omega^b$ traces corresponding to the two possible run modes in an optical
trap: i) bundle pointing to the negative $z$ direction and cell swimming upwards
with positive $z$ components for displacement and angular velocity, ii) bundle
pointing to the positive $z$ direction and cell swimming downwards with negative
$z$ components for displacement and angular velocity. A closer look to a zoomed 
10 s region also reveals the presence of tumble events that do not lead to a bundle 
reversal (dashed lines in Fig. \ref{fig:single}(b)). These events can be discriminated 
from bundle reversals (solid red lines in Fig. \ref{fig:single}(b)) since both $z$ and $\Omega^b$, after a $\sim$0.3 s transient where large fluctuations around zero are visible, both return to their initial value rather than changing sign.
These transient tumbling states however are not visible in the $z$
 and $\Omega^b$ histogram (Figure \ref{fig:single}(c)) for two 
reasons: i) the systems spends a negligible amount of time in a tumble compared to to the time spent 
in the two run states, ii) the values of $z$ and $\Omega^b$ are not well defined during a tumble.
The time trace of the cell polar angle $\theta$ reveals the presence of reorienting torques acting on the cell body during 
tumble and reversal events (Fig. \ref{fig:single}(c)). However, the restoring torque applied by the optical trap always prevents the cells from reorienting and constraints the polar angle below 5 degrees. 39 bacteria out of 53 displayed bundle reversals within our
observation time window. We counted a total of about 5400 tumble events, of
which 34\% led to a bundle reversal. This value is in between the previously
reported figures of 56\% \cite{Turner1995reversal} and 16\%
\cite{Chemla2009trapping}. For the particular cell shown in Fig \ref{fig:single}
we counted 99 tumble events of which 34 lead to a bundle reversal. The cell
spends about the same time in the two run states with an average run duration of
3 s. 
Fig. \ref{fig:tau} shows the cumulative run length distributions for the two states of 4 cells. Run length distributions display a high variability among different cells (different colours) but also between the two running states of an individual cell (solid and dashed lines).  The distribution of running times is exponential for run lengths shorter than ~ 5 s but shows a  systematic excess of long runs, as also found in \cite{Chemla2009trapping, shimizu}.
In particular, our data seem to show that the fat tail at long run lengths is more pronounced in one of the two running states.
Trap calibration is performed independently for each cell after a short exposure to a focused 532 nm beam that causes photodamage and stops motility \cite{opticution, neu, rasmu}.
Using our 3D imaging technique we can track the Brownian fluctuations of the cells position and orientation and reconstruct the optical potential assuming Boltzmann statistics (see Appendix).
Once the trap is calibrated we can directly convert $z$ displacements into
propelling forces. In particular we want to address the open question: is there
any change in the propelling force associated with rebundling of flagella on the
opposite cell pole? 
Before we go further with data analysis it is useful to consider a simple
hydrodynamic scheme for the swimming problem. Following \cite{Purcell1997} we
model a swimming cell as the combination of two rigid units: the cell body and
the flagellar bundle. The two units are rigidly connected but they are free to
rotate around a common axis. We can formulate the corresponding hydrodynamic
equations in terms of a resistance problem connecting force and torque
components along the swimming direction to the corresponding velocity and
angular frequency components:

\begin{eqnarray}
F^b&=&A^b U\label{first}\\
T^b&=&D^b\Omega^b\label{second}\\
F^f&=&A^f U+B^f\Omega^f\label{third}\\
T^f&=&B^f U+D^f\Omega^f\label{fourth}
\end{eqnarray}

\noindent where the superscripts $b$ and $f$ refer respectively to cell body and flagellar
bundle. The symbol $A$ represents the translational drag coefficient connecting the linear 
speed $U$ to the force $F$ that is needed to move a body in a purely translatory motion. 
Similarly $D$ connects angular speed to torque in pure rotations. For bodies that, 
like the helical bundle,  do not posses three mutually orthogonal planes of symmetry, a 
third coupling coefficient $B$ is required connecting the angular speed to the force that is 
required to prevent them from translating. 

For a trapped cell the swimming speed $U$ vanishes while the total external force is provided by the optical trap:
\begin{equation}
F^b+F^f=F_{trap}=B^f \Omega^f
\end{equation}
In other words the trapping force balances the thrust $F_{thrust}=-B^f\Omega^f$ from the bundle and prevents the cell from swimming.
We measure an average thrust of  0.14 pN which is lower than the previously 
reported values 0.57 pN and 0.4 pN respectively in \cite{Chattopadhyay2006trapping} 
and \cite{Goldstein2011}. This is consistent with the lower average swimming speed that we measured on our samples ($\sim$ 10 $\mu$m/s) compared to an almost double value ($\sim$ 20 $\mu$m/s) reported in \cite{Chattopadhyay2006trapping, Goldstein2011}. We believe this discrepancy could be due to differences in strain, buffer composition and sampling of bacterial tracks.

\begin{figure}[ht]
\hspace{-.3cm}
\includegraphics[width=0.48\textwidth]{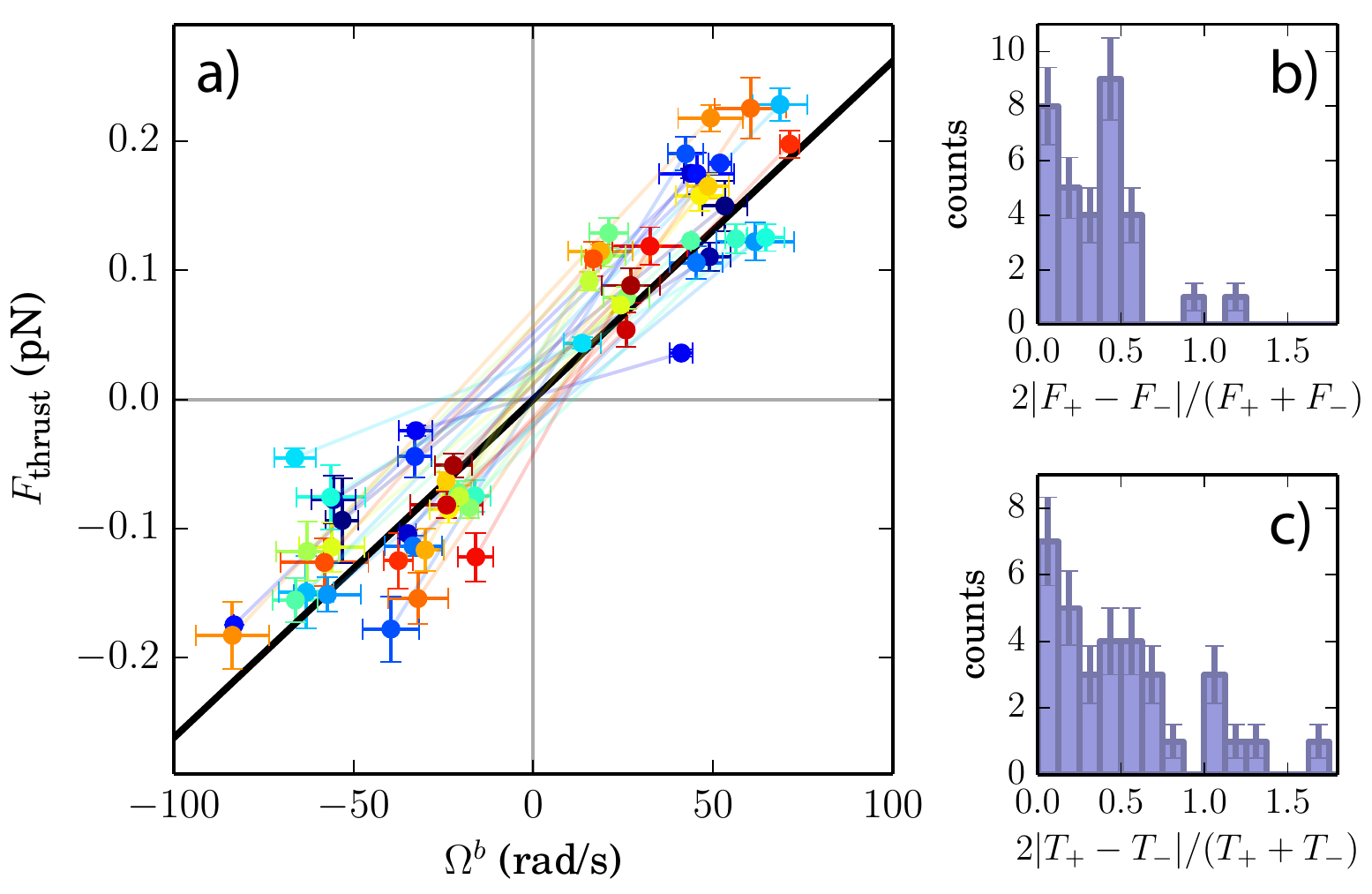}
\caption{(a) Thrust $F_\textrm{thrust}$ versus body angular frequency $\Omega^b$ for all the
bacteria undergoing bundle reversals. Different bacteria are shown in different
colours. Solid lines join the two run states for the same cell. Error bars
represent $\pm$ the standard deviation of $F_\textrm{thrust}$ and $\Omega^b$ between different
runs. (b) Histogram of the normalized force difference between the two run
states. (c) Histogram of the normalized torque difference between the two run
states.}
\label{fig:multi}
\end{figure}

In order to quantify the degree of variability of a generic positive quantity
$Q$ between the two run states, denoted by the subscript $+$ and $-$, we use the
normalised  difference $\Delta[Q]=2\langle |Q_+ - Q_-|/(Q_+ + Q_-)\rangle$ where
angle brackets denote averages over different cells. Fig. \ref{fig:multi}(b)
reports a histogram of the normalised force differences between the two states.
For the thrust we obtain $\Delta[F_{\textrm{thrust}}]=0.35$ that is much larger than the
average force fluctuations between different runs in the same direction (10\%).
In other words, run and tumble dynamics in \coli  is characterised by two
different and well defined run states corresponding to flagella bundling at one
of the two cell poles and resulting in propelling force differences that are on
average 35\% with tails extending over 100\%. This is in contrast with what
found in another peritrichously flagellated bacterium {\it B. subtilis}  where
very similar swimming speeds have been measured \cite{Kessler2006reversal}. The more symmetric
swimming behavior of {\it B. subtilis} could be related to a larger number of flagella
($\sim$ 30). As shown in Fig. \ref{fig:multi}(c), large differences are also
observed between the torque values in the two run states. Since $D^b$ does not
change upon bundle reversal we have, using (\ref{second}), $\Delta[T^b]=\Delta[\Omega^b]=0.45$ that is again much larger than torque
fluctuations between different runs pointing in the same direction (15\%). It is
interesting to note that, using the torque free condition $T^b=-T^f$ and  applying $\Delta[\cdot]$ to both sides of
(\ref{fourth}) we get  $\Delta[T^b]=\Delta[T^f]=\Delta[D^f]+\Delta[\Omega^f]$. If we assume,
as reported in \cite{Chemla2009trapping}, that the bundle rotation frequency
does not change much upon reversal then $\Delta[\Omega^f]\sim0$ and therefore
we obtain $\Delta[T^b]\sim\Delta[D^f]$. This seems to suggest that the observed
torque differences between the two run states have to be attributed mostly to
changes in bundle conformations leading to changes in the rotational drag $D^f$.
Large variations in $D^f$ could result from having one or more flagella rotating
out of the bundle.
From (\ref{second}) and (\ref{fourth}) we obtain that the propelling force
should be directly proportional to the measured body rotation frequency
$F_\textrm{thrust}=D^b\;B^f/D^f\;\Omega^b$. Since $D^b$ does not change upon bundle reversal,
any change in the ratio $F_\textrm{thrust}/\Omega^b$ has to be attributed to different bundle
conformations in the two states leading to a different $B^f/D^f$ ratio. Fig.
\ref{fig:multi}(a) reports a plot of $F_\textrm{thrust}$ versus $\Omega^b$ for all cells that
undergo at least one reversal. Solid lines connect points that refer to the same
cell. If the bundle conformation remained the same upon reversal, for each cell, the two points should 
lay on a line with slope $D^b B^f/D^f$ passing through the origin. This is clearly not the case
for the vast majority of cells.  From data shown in Fig \ref{fig:multi} we get
$\Delta[B^f/D^f]=\Delta[F_\textrm{thrust}/\Omega^b]=0.47$ indicating again a substantial
variation between bundle conformations in the two states. 
The solid black line in Fig. \ref{fig:multi}(a) represents the best fit through
the points whose slope ($0.0026\pm0.0001$ pN s/rad) provides the average value:

\begin{equation}
\left<\frac{F_\textrm{thrust}}{\Omega^b}\right>=\left<D^b\right>\left<\frac{B^f}{D^f}\right>\label
{five}
\end{equation} 

where we have assumed that body size and flagellar bundle conformation are
statistically independent.
The average value of $\langle D^b\rangle$ can be estimated approximating the
cell body with a prolate ellipsoid having the cell length $\ell$ as the major
axis and the cell thickness $a$ as the average minor axis, $D^b=\mu \pi
a^2\ell$ (where $\mu$ is the water viscosity). The average cell length and its corresponding standard error 
are obtained by 3D reconstructions ($\langle\ell\rangle=3.12\pm0.06\;\mu$m) while the width is independently measured on
fluorescently labeled cells ($\langle a\rangle=0.40\pm0.02\;\mu$m).
As a result we obtain an average value  for $\langle D^b\rangle=0.0063\pm0.0006$
pN $\mu$m s/rad. Using (\ref{five}) we find $\langle
B^f/D^f\rangle=0.41\pm0.04\;\mu\textrm{m}^{-1}$. Our value is significantly
lower then that found in \cite{Chattopadhyay2006trapping}, where the authors
obtained for the same ratio 1.13 $\mu$m$^{-1}$. Recent studies
\cite{Zhang2013helix} have shown that resistive force theories usually result in
an overestimation of the thrust $B^f$ of helical propellers while Lighthill's
slender body theory  \cite{light} provides a much more accurate result. In
particular, using average bundle parameters from \cite{Berg2000imaging} (length
7.3 $\mu$m, pitch 2.3 $\mu$m, diameter 0.36 $\mu$m) and assuming a bundle
radius of 0.02 $\mu$m we get from slender body theory a ratio $B^f/D^f=0.38$
that is in very good agreement with our measured value.

\section{Conclusions} 
We combined optical traps and 3D holographic microscopy to
investigate asymmetric features that arise in  \coli swimming as a consequence
of flagellar bundling at either pole of the cell. We found that two well
defined running states exist. These are characterised by large differences in 
propelling force, total applied torque and flagellar conformation. We also 
provide a direct measurement of the ratio $B^f/D^f$ which confirms predictions 
from slender body theory. Our technique can be easily extended to multiple cell 
systems allowing to study hydrodynamic couplings between individual swimming 
cells in well defined and reproducible configurations.

The research leading to these results has received funding from the European
Research Council under the European Union's Seventh Framework Programme
(FP7/2007-2013) / ERC grant agreement n$^\circ$ 307940. We also acknowledge
funding from  MIUR-FIRB project No. RBFR08WDBE.

\appendix
\section{APPENDIX: TRAP CALIBRATION}
\begin{figure}[h]
\includegraphics[width=0.45\textwidth]{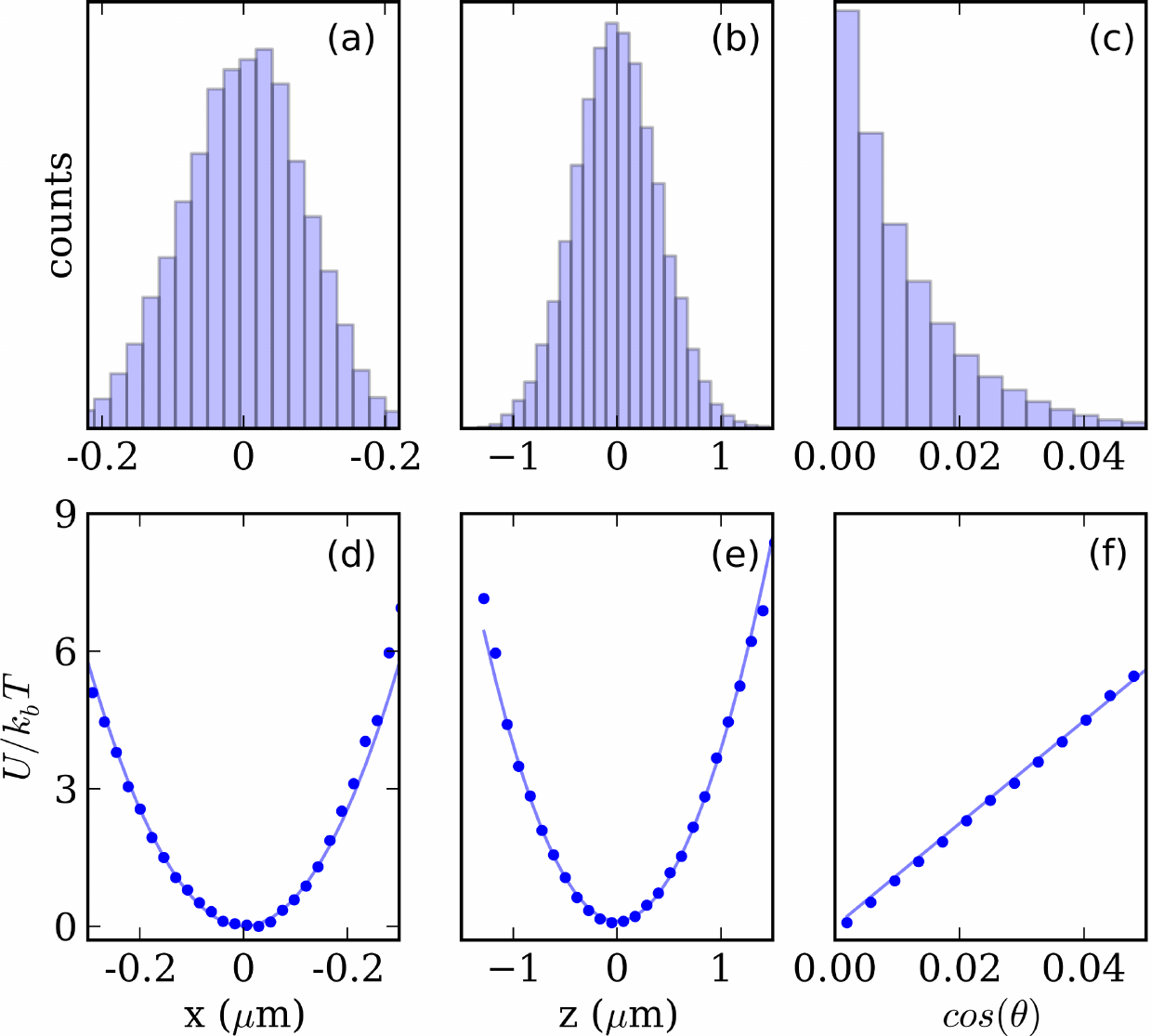}
\caption{Probability distribution (a-c) of $x$, $z$, $cos(\theta)$ of a trapped non-motile \textit{E. coli} cell. The corresponding potentials, measured in units of $k_{B}T$, are plotted as blue dots (d-f). The potentials are fitted with polynomials of respectively second, third, and first order that are plotted as blue lines.}
\label{fig:potential}
\end{figure}

Trap calibration is performed individually on each cell after a short exposure to 532 nm laser light that stops motility. The Brownian fluctuations of position $x,y,z$ and polar angle $\theta$ are recorded for 5 minutes in an optical trap of reduced intensity such that fluctuations in $z$ have a comparable amplitude to those observed with motile cells. For small fluctuations the optical energy is separable in the sum of independent contributions:

\begin{eqnarray}
\label{uncorr}
U(x,y,z,\theta)\simeq U_x(x)+U_y(y)+U_z(z)+U_\theta(\theta))
\end{eqnarray}
Thermal fluctuation will be then distributed according to Boltzmann statistics leading to a factorized probability density:
\begin{equation}
P(x,y,z,\theta)\propto e^{-\beta U_x(x)}e^{-\beta U_y(y)}e^{-\beta U_z(z)}e^{-\beta U_\theta(\theta)}
\end{equation}
with $\beta=1/k_B T$. This approximation is confirmed by the absence of correlation between the coordinates $(x,y,z,\theta)$.
The individual probability densities and the corresponding potentials are reported in Fig. \ref{fig:potential}.
We are interested here in converting $z$ displacements into optical forces. To this aim we fit $U_z(z)$ with a $3^{\textrm{rd}}$ order polynomial:
\begin{equation}
\label{poly}
 U(z)=\frac{1}{2}k z^2+\frac{1}{3}k^\prime z^3
\end{equation}
We find that $k^\prime/k$ is always of the order of $10^{-2}$ $\mu$m$^{-1}$. 
In our experiments the laser power is adjusted to the minimum value that is required for stable trapping of swimming cells. In this situation the cell body moves away from the trap's equilibrium position by 1.2 $\mu$m at most. Therefore, the third order term only affects force measurements at the level of a few percent that is below other sources of error.


\end{document}